\documentclass[sigconf]{acmart}

\AtBeginDocument{
  }

\setcopyright{cc}
\setcctype{by}
\copyrightyear{2026}
\acmYear{2026}
\acmDOI{10.1145/3805712.3809607}
\acmConference[SIGIR '26]{Proceedings of the 49th International ACM SIGIR Conference
on Research and Development in Information Retrieval}{July 2024, 2026}{Melbourne,
VIC, Australia}
\acmBooktitle{Proceedings of the 49th International ACM SIGIR Conference on Research
and Development in Information Retrieval (SIGIR '26), July 20--24, 2026, Melbourne, VIC,
Australia}
\acmISBN{979-8-4007-2599-9/2026/07}

\author{Yuan Li}
\affiliation{
  \institution{National University of Singapore}
  \country{Singapore}
}
\email{li.yuan@u.nus.edu}

\author{Jun Hu}
\authornote{Corresponding author.}
\affiliation{
  \institution{National University of Singapore}
  \country{Singapore}
}
\email{jun.hu@nus.edu.sg}

\author{Jiaxin Jiang}
\affiliation{
  \institution{National University of Singapore}
  \country{Singapore}
}
\email{jxjiang@nus.edu.sg}

\author{Bryan Hooi}
\affiliation{
  \institution{National University of Singapore}
  \country{Singapore}
}
\email{dcsbhk@nus.edu.sg}

\author{Bingsheng He}
\affiliation{
  \institution{National University of Singapore}
  \country{Singapore}
}
\email{dcsheb@nus.edu.sg}

\usepackage{multirow}
\usepackage{booktabs}
\usepackage{bm}
\usepackage{subfig}
\usepackage{pifont}
\usepackage{colortbl}
\usepackage[table]{xcolor}

\usepackage{amsfonts}
\usepackage{enumitem}
\usepackage{color}
\usepackage{graphics}
\usepackage{graphicx}
\usepackage[linesnumbered,ruled,vlined,algopart]{algorithm2e}

\newcommand{\cmark}{\ding{51}}  
\newcommand{\xmark}{\ding{55}}  
\newcommand{\ms}[2]{{#1{$\pm$#2}}}

\usepackage{balance}

\begin{document}

\renewcommand{\shortauthors}{Yuan Li, Jun Hu, Jiaxin Jiang, Bryan Hooi, and Bingsheng He}

\title[Robust Multimodal Recommendation via Graph Retrieval-Enhanced Modality Completion]{Robust Multimodal Recommendation via Graph Retrieval-Enhanced Modality Completion}

\begin{abstract}
Multimodal data plays a critical role in web-based recommendation systems, where information from diverse modalities such as vision and text enhances representation learning. However, real-world multimodal datasets often suffer from modality incompleteness due to sensor failures, annotation scarcity, or privacy constraints, which substantially degrade model performance and reliability. One effective solution to address this issue is modality completion, which reconstructs missing features to provide modality-complete graphs for downstream tasks. Given a query node with missing multimodal features, existing modality completion methods typically infer information from the node itself or its neighbors to reconstruct the missing modality. However, these methods may overlook semantically relevant context in the graph, which contains valuable cues that are non-trivial to capture through simple methods like neighborhood aggregation. In this work, we propose GRE-MC, a Graph Retrieval–Enhanced Modality Completion framework, to overcome these limitations. By introducing a modality-aware subgraph retrieval mechanism, GRE-MC selects semantically relevant subgraphs from the entire graph, providing richer contextual information for completing missing modalities. Subsequently, a graph transformer jointly encodes the query node and the retrieved subgraph via global attention to complete the missing features, while a learnable sparse-routing codebook regularizes latent embeddings into compact bases for improved robustness. Extensive experiments on multimodal recommendation benchmarks demonstrate that GRE-MC consistently outperforms state-of-the-art methods, validating the effectiveness of subgraph retrieval and joint-encoding graph transformer for robust modality completion.
\end{abstract}



\maketitle

\section{Introduction}

\begin{figure}[!t]
    \centering
    \subfloat[Training GNNs for recommendation with missing modalities, which degrade the performance of downstream recommendation models.]{
        \includegraphics[width=\linewidth]{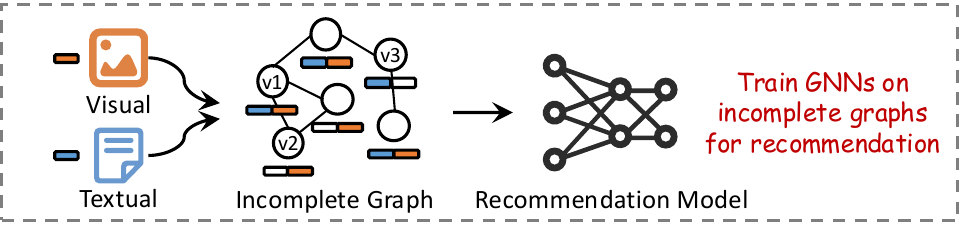}
        \label{fig:incomplete}
    }\\
    \subfloat[Training GNNs for recommendation with completed modalities, where enriched features provide informative signals for improved predictions.]{
        \includegraphics[width=\linewidth]{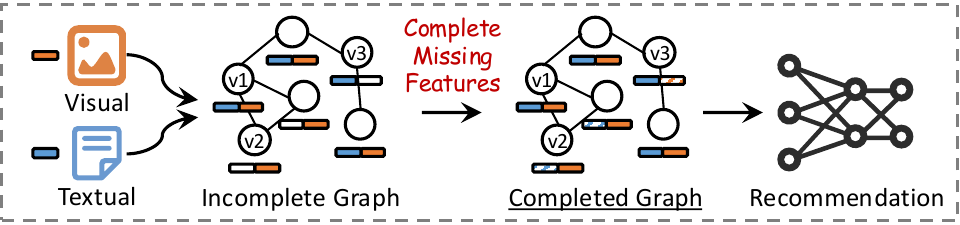}
        \label{fig:complete}
    }
\caption{Illustration of modality completion in recommendation tasks, where recovering missing features can improve downstream recommendation performance.}
    \label{fig:incomplete-vs-complete}
\end{figure}

Multimodal data, which integrates information from diverse modalities such as vision and text, has greatly advanced web-based recommendation applications~\citep{wei2019mmgcn, wei2020graph}.
Despite these advances, multimodal recommendation systems often face modality incompleteness caused by sensor malfunctions, limited annotations, or privacy constraints~\citep{li2025generating, bai2024multimodality, malitesta2024we}.
As illustrated in Figure~\ref{fig:incomplete}, a real-world graph may contain nodes with missing textual features (e.g., node $v_2$) or missing visual features (e.g., node $v_3$).
Most existing multimodal recommendation models~\citep{yu2025mind, cai2022adaptive, wei2019mmgcn} assume that all modalities are fully available (e.g., node $v_1$), without considering nodes with missing modalities during either training or evaluation, leading to substantial performance degradation~\citep{li2025generating, malitesta2024we} in real-world scenarios where certain modalities may be absent.

A range of solutions have been explored to mitigate the issue of missing modalities in web-based recommendation systems. Existing approaches focus on designing modality-robust models or leveraging auxiliary signals, often inspired by advances in multimodal learning such as learning modality-aligned representations~\citep{bai2024multimodality,tsai2019multimodal} and knowledge distillation to transfer information from complete-modality data~\citep{wang2020multimodal,wang2023learnable}.
While effective, such model-centric methods require modifying each downstream model to handle incomplete inputs, which introduces substantial implementation overhead and is impractical for large-scale deployments.
An alternative approach is \textbf{modality completion}, which aims to generate the missing modality data before passing inputs to the downstream model, as illustrated in Figure~\ref{fig:complete}. By reconstructing a complete set of modalities—using techniques such as graph-based feature propagation~\citep{malitesta2024we} or conditional generative models~\citep{li2025generating}—modality completion enables the use of existing multimodal models without architectural modifications.
In this work, we focus on modality completion for multimodal recommendation systems, aiming to handle missing modalities in practical recommendation scenarios.

Extensive prior work has advanced modality completion as a pre-processing strategy to reconstruct missing modalities before downstream learning, enabling seamless integration with existing multimodal architectures without the need for architectural modifications~\citep{bai2024multimodality,malitesta2024we,li2025generating}.
Broadly, these approaches can be categorized into two main paradigms:
(1) \textit{self-based methods}, such as MoDiCF~\citep{li2025generating} and LRMM~\citep{wang2018lrmm}, which exploit intra-instance correlations between available modalities (e.g., using textual features to reconstruct visual features). These models typically employ cross-modal attention or conditional generation networks to infer the missing modality from the available ones; and
(2) \textit{neighbor-based methods}, such as Imputed~\citep{malitesta2024we}, which leverage inter-instance dependencies—often captured through graph structures like co-purchase or co-click networks—to impute missing features by aggregating signals from related entities.
Both paradigms have demonstrated promising results in improving the completeness and quality of multimodal representations.

\begin{figure}[!t]
    \centering
    \includegraphics[width=\linewidth]{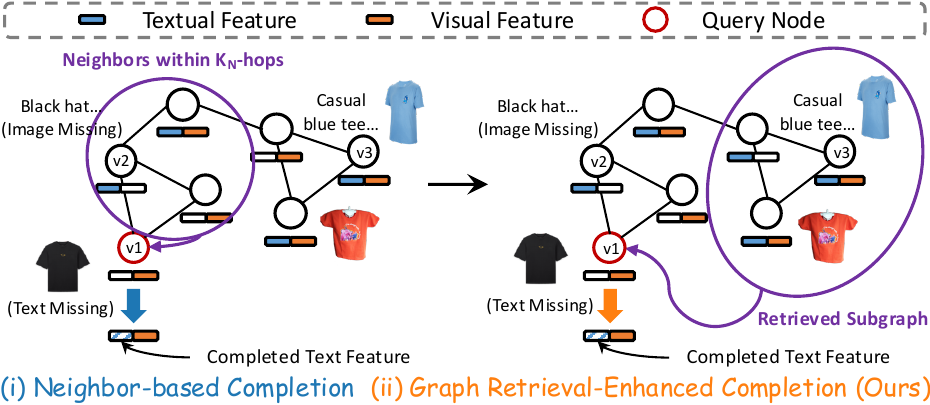}
    \caption{Motivation for graph retrieval–enhanced modality completion. (i) Neighbor-based completion may aggregate neighbors with less semantically relevant features. (ii) Graph retrieval–enhanced completion retrieves a distant yet semantically relevant subgraph.}
    \label{fig:neighbor-vs-retrieval}
\end{figure}

\begin{figure}[!t]
    \centering
    \includegraphics[width=0.7\linewidth]{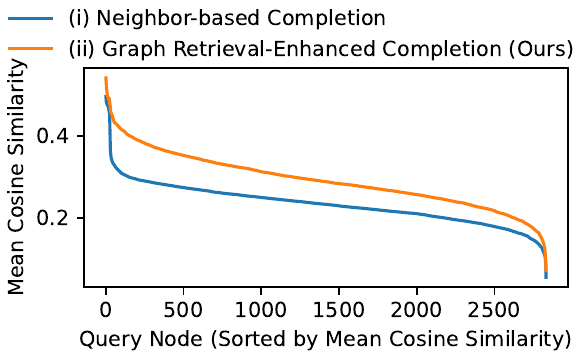}
    \caption{Comparison of subgraph relevance between neighborhoods and semantically retrieved subgraphs under missing-modality scenarios.}
    \label{fig:relevance}
\end{figure}

\textbf{Subgraph Retrieval for Modality Completion.} 
As depicted in Figure~\ref{fig:neighbor-vs-retrieval}(i), given a query node $v_1$ (a shirt) with missing textual features, neighbor-based methods consider its neighbors within $K_N$ hops as information relevant to $v_1$ and use them for modality completion. 
In this paper, we consider a new paradigm, subgraph retrieval, to extract information relevant to the query nodes. 
Specifically, as shown in Figure~\ref{fig:neighbor-vs-retrieval}(ii), our subgraph retrieval approach (detailed in Section~\ref{subsec:retrieval}) extracts a subgraph from the entire graph by leveraging both graph semantics derived from the available modalities of the query nodes and structural information, ensuring that the retrieved subgraph is semantically relevant.
Figure~\ref{fig:relevance} compares two strategies for obtaining nodes relevant to a query node: (i) its $K_N$-hop neighbor nodes (we set $K_N=2$ for illustration), and (ii) the nodes in its retrieved subgraph obtained through our subgraph retrieval approach.
Importantly, while the retrieval process relies on the available modalities of the query node, the relevance metric measures the average similarity between nodes obtained from each strategy and the missing modality of the query node. 
The results demonstrate that the retrieved subgraphs achieve significantly higher relevance scores than the neighborhood, indicating that our retrieval approach provides richer and more complementary contextual information for robust modality completion in scenarios with missing modalities.

\textbf{Graph Transformer for Joint Encoding.} 
In neighbor-based methods, the self-information of the query node and that of its neighboring nodes are jointly encoded to infer the missing modality. 
Typically, the query node and its neighbors are connected (e.g., including $K_N$-hop neighbors), and message passing is employed during encoding to mutually enrich their representations. 
However, in our framework, the neighboring nodes are replaced by a retrieved subgraph, where the query node may not be directly connected to any node within this subgraph, making conventional message passing between them infeasible. 
To address this, we introduce a graph transformer~\citep{yun2019graph,hu2026ntsformer} that enables the joint encoding of the query node and its retrieved subgraph through global attention, allowing information exchange regardless of graph connectivity. 
Global positional encodings are incorporated to provide global structural information for both the query node and the retrieved subgraph, thereby enabling the model to capture long-range dependencies even across disconnected components.
Furthermore, we introduce a learnable sparse-routing codebook that discretizes latent embeddings into a compact set of basis vectors, regularizing the representation space and enhancing robustness to incomplete modalities.

In this paper, we propose \textbf{GRE-MC}, a graph retrieval–enhanced framework for modality completion.
By retrieving semantically relevant subgraphs and encoding them jointly with the query node via a graph transformer, GRE-MC effectively leverages distant yet informative context to complete missing modalities.
Extensive experiments demonstrate that GRE-MC consistently outperforms existing approaches across multiple multimodal recommendation benchmarks.
The contributions are summarized as follows:
\begin{itemize}
\item We propose GRE-MC, a graph retrieval–enhanced framework for modality completion. By introducing a dual-stage subgraph retrieval mechanism, GRE-MC retrieves semantically relevant subgraphs for more informed completion.
\item We introduce a graph transformer that jointly encodes the query node and the retrieved subgraph to capture long-range dependencies. A sparse-routing codebook further discretizes latent representations into compact bases, improving robustness under incomplete modalities.
\item Extensive experiments on multimodal recommendation benchmarks demonstrate that GRE-MC enhances downstream performance, outperforming state-of-the-art (SOTA) methods under missing-modality scenarios.
\end{itemize}

\section{Preliminaries}

\textbf{Multimodal Recommendation} (MMRec) systems aim to enhance recommendation performance by incorporating diverse modalities, such as textual descriptions, images, acoustic features, and more~\citep{li2025generating, hu2025modality}. 
Each instance can be described by multiple modality-specific features, enabling richer user–item semantic modeling. 
Formally, let \( \mathcal{M} \) denote the set of available modalities, where each modality \( m \in \mathcal{M} \) is associated with a feature space. For each instance \( i \), its multimodal features are denoted as \(\{\mathbf{x}_i^{(m)}\}_{m \in \mathcal{M}}\). Unlike traditional recommenders that rely solely on user–item interactions, MMRec frameworks learn representations by integrating content from multiple modalities to better capture their characteristics.

\textbf{Graph-based Recommendation.}
Many SOTA recommendation systems model the recommendation problem using a user-item interaction graph \( \mathcal{G} = (\mathcal{V}, \mathcal{E}) \), where nodes \( \mathcal{V} \) represent users and items, and edges \( \mathcal{E} \) indicate observed interactions~\citep{hu2025modality}. 
We denote the number of nodes as $N$ and the number of edges as $E$, where the nodes consist of $N_U$ users and $N_I$ items.
These interactions are typically represented by an adjacency matrix \( \mathbf{A} \in \{0,1\}^{|\mathcal{V}| \times |\mathcal{V}|} \). Graph Neural Networks (GNNs)~\citep{kipf2017semisupervised, Hamilton2017GraphSAGE} and more graph learning approaches\citep{hu2026echoless, li2026dgp} are then employed to propagate information through the graph, modeling higher-order dependencies among users and items~\citep{wei2019mmgcn}. 

We further construct an item graph \( \mathcal{G}_I = (\mathcal{V}_I, \mathcal{E}_I) \) induced from the user-item interaction graph \( \mathcal{G} \).  
The node set \( \mathcal{V}_I \) consists of all items, i.e., \( |\mathcal{V}_I| = N_I \). 
Two items \( i, j \in \mathcal{V}_I \) are connected if they share at least one common user in \( \mathcal{G} \):
\begin{equation}
(i,j) \in \mathcal{E}_I \iff \exists \, u \in \mathcal{V}_U \ \text{s.t.} \ (u,i) \in \mathcal{E}, \, (u,j) \in \mathcal{E}
\end{equation}
where \( \mathcal{V}_U \) denotes the user set. 
This induced graph provides an item-centric view, enabling modality-complete items to propagate information to missing-modality nodes for more effective completion. 
Since item information is generally more publicly accessible, while user information tends to be more private, we focus on item completion in this work and hereafter use the symbol $\mathcal{G} = \mathcal{G}_I$ to denote the item graph, and similarly let $\mathcal{V} = \mathcal{V}_I$ and $\mathcal{E} = \mathcal{E}_I$.

\textbf{Missing Modalities and Completion.}
In real-world applications, item modalities are often incomplete due to privacy concerns, data sparsity, or acquisition costs~\citep{li2025generating}. This results in missing-modality scenarios, where some items are associated with only a subset of modalities. Formally, we define a modality indicator matrix \( \mathbf{E} \in \{0,1\}^{N_I \times |\mathcal{M}|} \), where \( \mathbf{E}_{i,m} = 1 \) if modality \( m \) is observed for item \( i \), and \( \mathbf{E}_{i,m} = 0 \) otherwise. Given the observed modalities, the item’s available feature set is \(\{\mathbf{x}_i^{(m)} : \mathbf{E}_{i,m} = 1\}\).
The objective of modality completion is to learn a conditional completion function:
\begin{equation}
\mathcal{C} \colon (\mathcal{G}, \{\mathbf{x}_i^{(m)} : \mathbf{E}_{i,m}=1\}, \mathbf{E}_i) 
\longrightarrow \{\hat{\mathbf{x}}_i^{(m)} : \mathbf{E}_{i,m}=0\}
\end{equation}

\section{Related Work}

\subsection{Multimodal Recommendation Models}
Multimodal recommendation enhances user or item embeddings with auxiliary modalities (e.g., text, images) to complement interaction signals. 
Traditional baselines such as BPR~\citep{rendle2009bpr} and LightGCN~\citep{he2020lightgcn} rely solely on interaction graphs and serve as unimodal benchmarks. 
More recent approaches incorporate multimodal information: 
SLMRec~\citep{tao2022self} applies self-supervised tasks to align and denoise multimodal views; 
BM3~\citep{zhou2023bootstrap} uses contrastive alignment across modalities. 
Despite their strong performance under complete modality settings, these models largely assume that all modalities are observed. As a result, they experience significant degradation in real-world scenarios where item modalities are missing or incomplete, motivating research into modality-aware completion. 

\subsection{Self-Based Modality Completion}
Previous work has explored self-based modality completion, which leverages a node’s own observed modalities to infer its missing ones, either by imputing features or by learning modality-invariant representations~\citep{wu2024deep,zhang2024multimodal}. For example, SDR-GNN~\citep{fu2024sdr} reconstructs missing signals in the spectral domain, while MMImputeDiff~\citep{li2024mmimputediff} employs diffusion models to synthesize missing imaging modalities. Although effective, these approaches mainly operate at the instance level and do not leverage structural signals available in recommendation graphs. As a result, they struggle when items lack strong intra-instance modality correlations or when the observed modalities provide weak guidance. Our work addresses this limitation by augmenting self-based completion with retrieval from semantically related subgraphs, enabling completion that is both structure-aware and robust under modality missingness.

\subsection{Neighbor-Based Modality Completion}
GNNs are widely used to model user–item relations in multimodal recommendation, as they capture higher-order dependencies through message passing~\citep{zhang2024dglf,chen2024joint,xing2024emo}. 
However, prior graph-based modality completion methods largely rely on neighbor aggregation~\citep{malitesta2024we}, assuming that neighbors provide informative signals for reconstructing missing modalities. 
As illustrated in Figure~\ref{fig:neighbor-vs-retrieval}, this assumption can be fragile: $K_N$-hop neighbors may be incomplete or semantically irrelevant, leading to error propagation and poor completion quality. Our approach overcomes this shortcoming by introducing a graph retrieval mechanism that adaptively selects globally relevant subgraphs instead of restricting completion to $K_N$-hop neighborhoods. 
This retrieval step ensures that modality completion is guided by higher-quality, semantically relevant contexts even when neighbor nodes are unreliable.

In summary, existing modality completion approaches either focus on self-based feature recovery without leveraging structural signals or rely heavily on neighborhood aggregation in graphs, which is often brittle in noisy contexts.
These limitations motivate a new paradigm that jointly exploits global graph semantics and structural retrieval for robust modality completion.

\section{Methodology}

In this section, we introduce a graph retrieval-enhanced modality completion framework designed to address the challenge of missing modalities in recommendation systems. 

\begin{figure*}[!t]
    \centering
    \includegraphics[width=\linewidth]{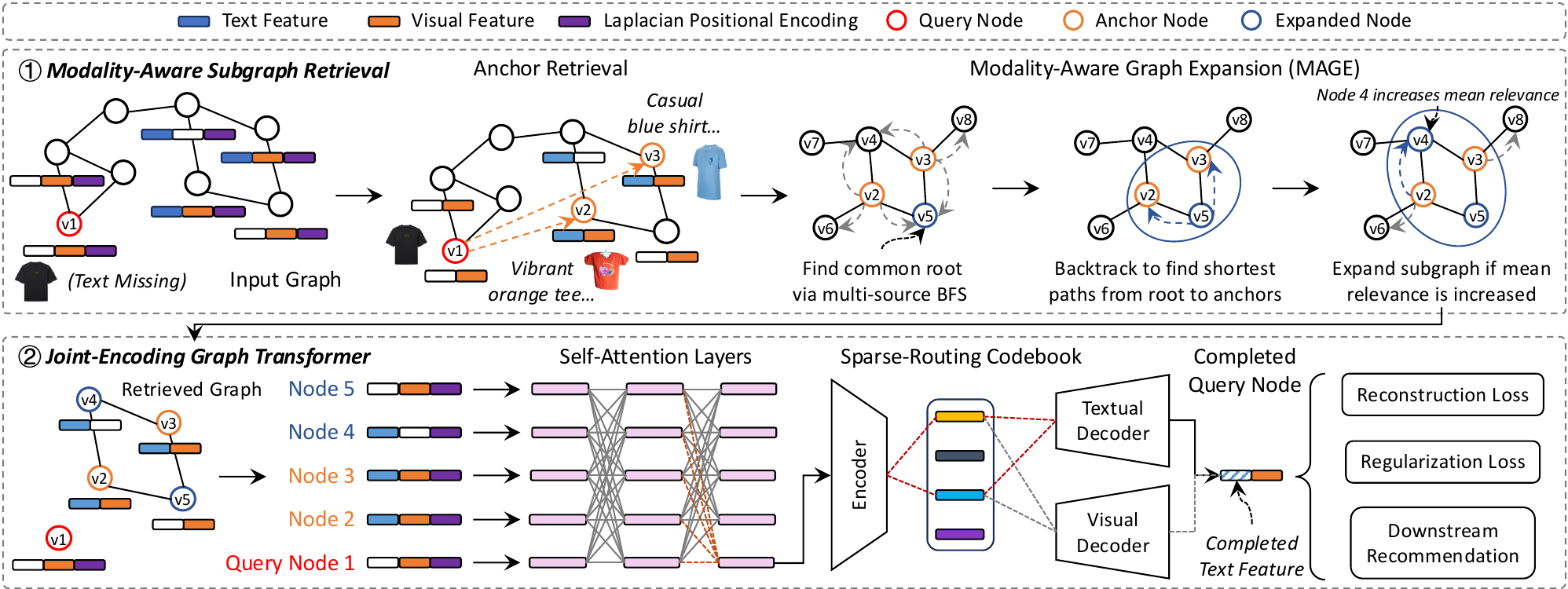}
    \caption{Overview of GRE-MC for graph retrieval-enhanced modality completion. The framework consists of two modules: (1) Modality-aware subgraph retrieval identifies semantically relevant anchor nodes and expands them into a structurally coherent subgraph via modality-aware graph expansion, integrating both semantic and structural signals; (2) Joint-encoding graph transformer encodes the retrieved subgraph to capture higher-order dependencies and cross-modal interactions, followed by sparse-routing codebook discretization to produce robust and compact modality representations.}
    \label{fig:overview}
\end{figure*}

\subsection{Overview}\label{subsec:overview}
Unlike prior methods that either (i) infer missing modalities from self-information or (ii) rely on neighbor aggregation, we propose GRE-MC, a graph retrieval–enhanced framework for modality completion in recommendation systems. As depicted in Figure~\ref{fig:overview}, it consists of two main components that jointly enable graph semantic- and structure-aware modality completion:

\begin{itemize}
\item \textbf{Modality-Aware Subgraph Retrieval} identifies a set of semantically related nodes based on available modalities and expands them into a structurally coherent subgraph through modality-aware graph expansion.
\item \textbf{Joint-Encoding Graph Transformer} encodes the retrieved subgraph with Laplacian positional encodings to capture higher-order dependencies and cross-modal interactions, and employs a sparse-routing codebook to discretize latent representations for robust modality completion.
\end{itemize}

\subsection{Modality-Aware Subgraph Retrieval}\label{subsec:retrieval}

This component first performs anchor set retrieval to identify a set of semantically relevant nodes, i.e., anchors, and then applies modality-aware graph expansion to integrate these anchors into a structurally coherent subgraph that captures both graph semantic and structural signals.

\textbf{Anchor Set Retrieval.} 
To retrieve informative context for reliable completion, GRE-MC leverages the entire graph, which contains abundant semantically relevant nodes, i.e., anchors. By retrieving such anchors, the model injects external signals beyond the query node’s self-information or $K_N$-hop neighborhood~\citep{mavromatis2025gnn, li2025rgl}. Specifically, consider query node $i$ and its available modality \( m \) such that \(\mathbf{E}_{i,m} = 1\), we perform nearest-neighbor search~\citep{douze2024faiss,indyk1998approximate} using modality $m$ to retrieve a set of anchors:
\begin{equation}
\mathcal{A}_i^{(m)} = \text{TopK}\!\left(\mathbf{x}_i^{(m)}, \{\mathbf{x}_j^{(m)} \mid \mathbf{E}_{j,m} = 1\}\right)
\end{equation}
The retrieved anchors serve as context that shares semantic similarity with the query node’s observed modalities, ensuring that the signals guiding completion are content-aware rather than purely structural. 
However, anchor nodes may be scattered across the graph without forming cohesive structures, which fail to express the underlying structural signals. 

\textbf{Modality-Aware Graph Expansion.}
Drawing inspiration from lowest common ancestors~\citep{yu2010keyword}, we enrich the semantic anchors with structural context by expanding them into an Anchor-Connecting Subgraph (ACS in Algorithm~\ref{alg:acs}) via multi-source breadth-first search:
\begin{equation}
S_0=\text{ACS}(\mathcal{G},\, \mathcal{A}_i^{(m)})
\end{equation}
The initial set $S_0$ connects the anchors through shortest paths. This expansion integrates higher-order structure with semantically retrieved nodes. 
We then refine this subgraph using a modality-aware iterative approach that maximizes the overall semantic correlation between the query node and the connected subgraph.

For any node \(v\), we define a pair-wise relevance score that aggregates cosine similarities over the modalities jointly observed between the query node \(i\) and node \(v\):
\begin{equation}\label{eq:r_piecewise}
\small
r(i, v) =
\begin{cases}
\dfrac{
\sum_{m \in \mathcal{M}} 
\mathbf{E}_{i,m}\,\mathbf{E}_{v,m}\,
\cos\big(\mathbf{x}_i^{(m)},\,\mathbf{x}_v^{(m)}\big)
}{
\sum_{m \in \mathcal{M}} 
\mathbf{E}_{i,m}\,\mathbf{E}_{v,m}
},
& \text{if } \sum_{m \in \mathcal{M}} 
\mathbf{E}_{i,m}\,\mathbf{E}_{v,m} > 0 \\
-\infty, & \text{otherwise}
\end{cases}
\end{equation}
Consequently, we define the mean relevance score between query node $i$ and node set $S$ as $\Phi(i,S)=\frac{1}{|S|}\sum_{v\in S} r(i, v)$.

Subject to the constraints that $S$ remains connected and contains all anchors, we refine the subgraph by improving its mean relevance. Starting from $S_0$, we iteratively apply two local moves to improve $\Phi(i,S)$: (i) add a boundary neighbor $c\notin S$ if $\Phi(i,S\cup\{c\})>\Phi(i,S)$; (ii) remove a non-bridging node $u\in S\setminus\mathcal{A}_i^{(m)}$ if $\Phi(i, S\setminus\{u\})>\Phi(i,S)$. 
At each step, we take the better of the two improvements and stop when neither addition nor removal can increase $\Phi$. 
This iterative process couples semantic relevance with connectivity: low-similarity non-bridging nodes are pruned, while highly relevant neighbors are greedily absorbed, yielding a semantically informative subgraph for downstream encoding. 
Combining the steps above, we obtain a subgraph via modality-aware graph expansion (MAGE in Algorithm~\ref{alg:mage}) with up to $T$ iterations:
\begin{equation}
S=\text{MAGE}(\mathcal{G},\, \mathcal{A}_i^{(m)}, T)
\end{equation}

Overall, our modality-aware subgraph retrieval integrates global graph semantics and structure, providing a robust foundation for subsequent graph encoding and modality completion. 

\subsection{Joint-Encoding Graph Transformer}
\label{subsec:generation}

After subgraph retrieval, we obtain a semantically relevant and structurally coherent subgraph. The next step is to transform this context into reliable representations for modality completion. We achieve this through a combination of graph transformer encoding and sparse-routing codebook discretization.

\textbf{Graph Transformer for Joint Encoding.}\label{sec:gt}
To capture rich dependencies within the retrieved subgraph, we represent each node as a token equipped with Laplacian positional encodings~\citep{maskey2022generalized} and process them with a Transformer~\citep{vaswani2017attention}. Compared with GCNs, this design preserves long-range, position-aware signals for the query node and the potentially disconnected retrieved subgraph.  
Formally, for adjacency matrix \(\mathbf{A}\) and degree matrix \(\mathbf{D}\), the normalized Laplacian matrix is defined as $\mathbf{L} = \mathbf{I} - \mathbf{D}^{-1/2}\mathbf{A}\mathbf{D}^{-1/2}$.
Let \(\mathbf{U}_k=[\mathbf{u}_1,\dots,\mathbf{u}_k]\) denote the bottom \(k\) nontrivial eigenvectors. The Laplacian positional encoding for node \(j\) is \(\mathbf{p}_j = [u_{j1},u_{j2},\dots,u_{jk}]\).  
We process the input embeddings for all nodes in the retrieved subgraph \( S \) 
and the query node \( i \) as:
\begin{equation}
\mathbf{h}_v^{(0)} =
\text{MLP}\!\left(\left[\oplus_{m \in \mathcal{M}} \tilde{\mathbf{x}}_v^{(m)};\mathbf{p}_v\right]\right),
\quad v \in S \cup \{i\}
\end{equation}
where \(\oplus\) denotes concatenation, and we handle missing modalities by defining 
\(\tilde{\mathbf{x}}_v^{(m)} = \mathbf{x}_v^{(m)}\) if \(\mathbf{E}_{v,m}=1\) and 
\(\tilde{\mathbf{x}}_v^{(m)}=\mathbf{0}\) otherwise.
Then, we adopt an $L$-layer Transformer to model long-range dependencies across the query node and the retrieved subgraph, allowing the query node to attend to the retrieved nodes and build context-aware representations:
\begin{equation}
\mathbf{H}^{(l)} = \text{TransformerLayer}(\mathbf{H}^{(l-1)}), \quad l=1,\dots,L
\end{equation}
where $\mathbf{H}^{(L)} = [\mathbf{h}_1^{(L)}, \dots, \mathbf{h}_n^{(L)}]$ and $n = |S| + 1$ denotes the total number of nodes, including the query node $i$ and the retrieved subgraph $S$.
Finally, the query embedding is computed by aggregating contextualized node representations with attention scores:  
\begin{equation}
\mathbf{z}_i = \sum_{j=1}^{n}\alpha_{ij}\,\mathbf{h}_j^{(L)},
\; \text{where}\; \alpha_{ij} = \frac{\exp\!\left((\mathbf{q}_i\cdot \mathbf{k}_j)/\sqrt{d_k}\right)}{\sum_{j'=1}^n \exp\!\left((\mathbf{q}_i\cdot \mathbf{k}_{j'})/\sqrt{d_k}\right)}
\end{equation}

This encoding scheme integrates structural positions and long-range dependencies, enabling the query embedding to capture rich contextual information from the retrieved subgraph.

\textbf{Sparse-Routing Codebook.}
To discretize the latent features for robust completion, we adopt a sparse-routing codebook mechanism with Gumbel--Softmax reparameterization~\citep{jang2017categorical,maddison2016concrete}. 
Given query embedding $\mathbf{z}_i$, we compute the routing weights:
\begin{equation}
\mathbf{g}_i = \mathrm{Softmax}\!\left((\boldsymbol{W}\mathbf{z}_i + \boldsymbol{\epsilon}) / \tau\right)
\label{eq:gumbel}
\end{equation}
where $\boldsymbol{W}$ is a learnable projection, $\tau$ is the temperature, and 
$\boldsymbol{\epsilon} = -\log(-\log \mathbf{u})$ with $\mathbf{u}\in\mathbb{R}^{C}$ and $u {\sim}\mathrm{Uniform}(0,1)$ 
denote the standard elementwise Gumbel perturbations used during training to enable differentiable sampling. 
Each $\mathbf{g}_i$ softly selects a few entries from the codebook $\{\mathbf{c}_e\}_{e=1}^C$, yielding a weighted sum of the top-$P$ discretized representations:
\begin{equation}
\mathbf{q}_i = \sum_{e \in \mathrm{TopP}(\mathbf{g}_i)} \mathbf{g}_{i,e}\,\mathbf{c}_e
\label{eq:q}
\end{equation} 
During inference, we set $\boldsymbol{\epsilon}=\mathbf{0}$ for deterministic routing. 
This discretized and sparse formulation promotes the reuse of prototypical factors and enables robust cross-modal reconstruction:
\begin{equation}
\hat{\mathbf{x}}_i^{(m)} = \mathrm{Decode}\big(\mathbf{q}_i, m\big) = \mathrm{MLP}^{(m)}(\mathbf{q}_i)
\label{eq:decode}
\end{equation}

\textbf{Codebook Regularization.}  
To avoid code collapse and encourage balanced codebook usage, we adopt regularization terms inspired by mixture-of-experts techniques~\citep{shazeer2017outrageously,fedus2022switch}, which employ auxiliary losses to promote balanced code assignments.  
Given a batch of size \(B\), and letting \(\mathbf{g}_i \in \mathbb{R}^C\) denote the soft routing probability over the \(C\) codebook entries for the \(i\)-th sample, the mean soft distribution \(\bar{\mathbf{g}}=\tfrac{1}{B}\sum_{i=1}^{B}\mathbf{g}_i\) and the uniform distribution \(\mathbf{u}=\tfrac{1}{C}\mathbf{1}\),  
we define a uniform-usage regularization term to encourage a globally balanced routing probability: 
\begin{equation}
\mathcal{L}_{\mathrm{usage}}
\;=\; \mathrm{KL}\!\left(\bar{\mathbf{g}} \,\middle\|\, \mathbf{u}\right)
\;=\; \sum_{e=1}^{C}\bar{g}_e \log \frac{\bar{g}_e}{1/C}
\label{eq:usage}
\end{equation}
where \(\bar{g}_e\) denotes the \(e\)-th element of \(\bar{\mathbf{g}}\).

Furthermore, after applying $\mathrm{TopP}(\mathbf{g}_i)$ routing with capacity control, we obtain discrete assignments \(\hat{\mathbf{g}}_i \in \{0,1\}^C\).  
We compute the empirical load of each code as \(\bar{\hat{\mathbf{g}}} = \tfrac{1}{B}\sum_{i=1}^{B} \hat{\mathbf{g}}_i\),  
and introduce a load-balancing loss to encourage uniform discrete assignments:  
\begin{equation}
\mathcal{L}_{\mathrm{load}}
\;=\; C \sum_{e=1}^{C}\bar{\hat{g}}_e^2
\label{eq:load}
\end{equation}

\textbf{Final Objective.}  
The overall training objective combines the reconstruction loss with two codebook regularization terms:  
\begin{equation}
\begin{gathered}
\mathcal{L}
=\mathcal{L}_{\mathrm{recon}}
+\lambda_{\mathrm{usage}}\mathcal{L}_{\mathrm{usage}}
+\lambda_{\mathrm{load}}\mathcal{L}_{\mathrm{load}}\\
\text{where}\quad
\mathcal{L}_{\mathrm{recon}}
=\sum_{m\in \mathcal{M}}\sum_{i}\big\|\hat{\mathbf{x}}_i^{(m)}-\mathbf{x}_i^{(m)}\big\|^2
\end{gathered}
\end{equation}
This joint objective enforces both soft uniform usage and hard load balancing across the codebook, ensuring stable training and reducing the risk of underutilized or collapsed codes.

Overall, by retrieving a semantically and structurally relevant subgraph, leveraging graph transformer for joint encoding, and employing a discrete sparse-routing codebook, GRE-MC produces robust completions that are modality-aware and structurally grounded.

\begin{algorithm}[!pt]
\caption{Anchor-Connecting Subgraph (ACS)}
\label{alg:acs}
\DontPrintSemicolon
\SetKwInOut{Input}{Input}
\SetKwInOut{Output}{Output}
\SetKw{KwRet}{return}
\Input{Graph $\mathcal{G}=(\mathcal{V},\mathcal{E})$, seed nodes $\mathcal{S}\subseteq\mathcal{V}$}
\Output{Connected subgraph $S$}
\BlankLine
\ForEach{$v\in\mathcal{V}$}{
  $\mathrm{dist}[v]\leftarrow +\infty$; 
  $\mathrm{parent}[v]\leftarrow -1$; \;
  $\mathrm{mask}[v]\leftarrow 0$ \tcp{bitmask of seed reachability}
}
$\text{index}:\mathcal{S}\to\{0,\dots,|\mathcal{S}|-1\}$ \tcp{encode seeds as bits}
$\text{full\_mask}\leftarrow (1\ll |\mathcal{S}|)-1$ \tcp{reach all seeds}
$Q\leftarrow \emptyset$ \tcp{FIFO queue}
\ForEach{$s\in\mathcal{S}$}{
  $i\leftarrow \text{index}(s)$; 
  $\mathrm{dist}[s]\leftarrow 0$; 
  $\mathrm{mask}[s]\leftarrow \mathrm{mask}[s]\ \mathbf{or}\ (1\ll i)$; 
  $Q.\mathrm{push}(s)$ \tcp{init}
}
$\mathrm{root}\leftarrow -1$ \;
\While{$Q$ not empty}{
  $u\leftarrow Q.\mathrm{pop}()$ \;
  \If{$\mathrm{mask}[u]=\text{full\_mask}$}{
    $\mathrm{root}\leftarrow u$; \textbf{break} \tcp{all waves collided here}
  }
  \ForEach{$w: (u,w)\in\mathcal{E}$}{
    $d\leftarrow \mathrm{dist}[u]+1$ \;
    \uIf{$d < \mathrm{dist}[w]$}{
      $\mathrm{dist}[w]\leftarrow d$; \;
      $\mathrm{parent}[w]\leftarrow u$; \;
      $\mathrm{mask}[w]\leftarrow \mathrm{mask}[u]$; \;
      $Q.\mathrm{push}(w)$ \;
    }\uElseIf{$d = \mathrm{dist}[w]$}{
      $m \leftarrow \mathrm{mask}[w]\ \mathbf{or}\ \mathrm{mask}[u]$ \;
      \If{$m \ne \mathrm{mask}[w]$}{
        $\mathrm{mask}[w]\leftarrow m$; \;
        $Q.\mathrm{push}(w)$ \tcp{merge reachability}
      }
    }
  }
}
\uIf{$\mathrm{root} = -1$}{
  \KwRet{$\bigcup_{s\in\mathcal{S}} \{v\mid \mathrm{dist}_s(v)\le d\}$}
}
\ForEach{$v\in\mathcal{V}$}{$\mathrm{parent}[v]\leftarrow -1$; $\mathrm{visited}[v]\leftarrow \text{false}$ \tcp{re-rooted BFS}}
$B\leftarrow \emptyset$; $B.\mathrm{push}(\mathrm{root})$; $\mathrm{visited}[\mathrm{root}]\leftarrow \text{true}$ \;
\While{$B$ not empty}{
  $u\leftarrow B.\mathrm{pop}()$ \;
  \ForEach{$w:(u,w)\in\mathcal{E}$}{
    \If{$\mathrm{visited}[w]=\text{false}$}{
      $\mathrm{visited}[w]\leftarrow \text{true}$; 
      $\mathrm{parent}[w]\leftarrow u$; 
      $B.\mathrm{push}(w)$ 
    }
  }
}
$S \leftarrow \emptyset$ \tcp{backtrack all seeds to root}
\ForEach{$s\in\mathcal{S}$}{
  $x\leftarrow s$ \;
  \While{$x\neq -1$}{
    $S \leftarrow S \cup \{x\}$ \;
    \If{$x=\mathrm{root}$}{\textbf{break}}
    $x \leftarrow \mathrm{parent}[x]$ \;
  }
}
\KwRet{$S$}
\end{algorithm}

\begin{algorithm}[!pt]
\caption{Modality-Aware Graph Expansion (MAGE)}
\label{alg:mage}
\DontPrintSemicolon
\SetKwInOut{Input}{Input}
\SetKwInOut{Output}{Output}
\SetKw{KwRet}{return}
\Input{Graph $\mathcal{G}=(\mathcal{V},\mathcal{E})$, anchors $\mathcal{A}_i^{(m)}$ for query node $i$ and modality $m$, iteration cap $T$, modality indicator $\mathbf{E}\in\{0,1\}^{|\mathcal{V}|\times|\mathcal{M}|}$}
\Output{Retrieved subgraph $S$}

$S \leftarrow \mathrm{ACS}(\mathcal{G},\, \mathcal{A}_i^{(m)})$ \tcp{Anchor-Connecting Subgraph}

\ForEach{$v \in \mathcal{V}$}{
  $d_{iv} \leftarrow \sum_{m\in \mathcal{M}} \mathbf{E}_{i,m}\,\mathbf{E}_{v,m}$ \tcp{jointly observed}
  \uIf{$d_{iv} = 0$}{
    $r(i,v) \leftarrow -\infty$ \;
  }\Else{
    $r(i,v) \leftarrow \dfrac{\sum_{m\in \mathcal{M}} \mathbf{E}_{i,m}\,\mathbf{E}_{v,m}\, \cos\!\big(\mathbf{x}_i^{(m)}, \mathbf{x}_v^{(m)}\big)}{d_{iv}}$ \;
  }
}

$\mu \leftarrow \dfrac{1}{|S|}\sum_{v\in S} r(i,v)$ \tcp{current mean relevance}

\Repeat{no improvement or iteration cap $T$ reached}{
  $\mathcal{N}(S) \leftarrow \{\, c \notin S \mid \exists\, u\in S,\ (u,c)\in \mathcal{E} \,\}$ \; 
  $c^\star \leftarrow \arg\max\limits_{c\in \mathcal{N}(S)} \Big(\dfrac{|S|\,\mu + r(i,c)}{|S|+1} - \mu\Big)$ \;
  $\Delta_{+} \leftarrow \dfrac{|S|\,\mu + r(i,c^\star)}{|S|+1} - \mu$ \;

  $\mathcal{R}(S) \leftarrow \{\, u \in S \setminus \mathcal{A}_i^{(m)} \mid S\setminus\{u\} \text{ is connected} \,\}$ \;
  \uIf{$\mathcal{R}(S)\neq \emptyset$}{
    $u^\star \leftarrow \arg\max\limits_{u\in \mathcal{R}(S)} \Big(\dfrac{|S|\,\mu - r(i,u)}{|S|-1} - \mu\Big)$ \;
    $\Delta_{-} \leftarrow \dfrac{|S|\,\mu - r(i,u^\star)}{|S|-1} - \mu$ \;
  }\Else{
    $\Delta_{-} \leftarrow -\infty$ \;
  }

  \uIf{$\max(\Delta_{+},\Delta_{-}) > 0$}{
    \uIf{$\Delta_{+} \ge \Delta_{-}$}{
      $S \leftarrow S \cup \{c^\star\}$ \;
    }\Else{
      $S \leftarrow S \setminus \{u^\star\}$ \;
    }
    $\mu \leftarrow \dfrac{1}{|S|}\sum_{v\in S} r(i,v)$ \;
  }\Else{
    \textbf{stop} \tcp{no improvement}
  }
}
\KwRet{$S$} \;
\end{algorithm}

\subsection{Complexity Analysis}\label{sec:complexity}

\textbf{Time Complexity.}
In practical recommendation graphs, effectively managing the computational footprint is crucial for scalable learning \cite{wu2025graphhash, li2025adapting}. We provide the upper bounds for a single query node $i$. Without loss of generality, we assume $|\mathcal{M}|=2$, $\mathbf{E}_{i,0}=1$, $\mathbf{E}_{i,1}=0$, and a uniform feature and hidden dimension $d$, so node $i$'s first modality is observed while the second is missing. Let $D=\mathrm{diam}(\mathcal G)$ be the graph diameter. 
First, the anchor set retrieval step uses nearest neighbor search to retrieve the top $K$ anchors for the observed modality, which costs $O\big(Nd+N\log K\big)$ and is thus dominated by $O(Nd)$ when $K \ll N$.
Then, the ACS graph expansion (Algorithm~\ref{alg:acs}) consists of root finding and graph construction. Finding a root that connects all anchors costs $O(E)$, while building the union of at most $K$ shortest paths of length at most $D$ costs $O(KD)$. 
The subsequent MAGE step (Algorithm~\ref{alg:mage}) with an iteration cap $T$ takes $O(TKD)$.
We note that the expansion frontier size can increase rapidly with large node degrees; however, this effect is not severe in practice, as we observe that the maximum node degree in the evaluated graphs is typically much smaller than the total number of nodes.
Then, the Laplacian positional encoding with $k$ eigenvectors costs $O(kKD)$.
A graph transformer with token length $O(KD)$ and $L$ layers takes $O\big(L\,(KD)^2 d \;+\; L\,(KD)\,d^2\big)$.
Finally, the codebook router with size $C$ costs $O(Cd)$, the top-$P$ selection takes $O(C+P\log P) $, and $L$-layer MLP decoder costs $O(Ld^2)$. 
In summary, the overall time complexity is
$
O(
Nd + E + KD(1+k+T)
+ L\big((KD)^2 d + (KD)d^2\big)
+ Cd + Ld^2
)
$.
In real-world recommendation graphs, most of the above factors are constrained (typically $D,K,k,T,C,L<10$ and $d \le 512$). 
Although the worst-case BFS cost can reach $O(E)$, in practice, the expansion usually terminates early once anchors meet within a few hops. 

\textbf{Space Complexity.}
Following the analysis above, the memory consumption involves storing the retrieved subgraph of size $O(KD)$, the Laplacian positional encodings $O(kKD)$, and the Transformer requires $O((KD)d+L(KD)^2)$ due to token features and attention maps. The sparse-routing codebook and router add $O(Cd)$ parameters and $O(C)$ temporary storage, while the decoder adds $O(Ld^2)$ parameters but only $O(d)$ activations per query. 
Overall, the space complexity is 
$
O(L(KD)^2 + KDd + kKD + C + Ld^2 + Cd)
$.
In practical recommendation graphs, both the subgraph size and the parameter size remain relatively small, so the overall memory consumption is within manageable limits.

\section{Experiments}
In this section, we evaluate the effectiveness and efficiency of GRE-MC on public benchmarks under missing-modality settings. 

\subsection{Experimental Setup}
We conduct experiments on three public benchmarks using a Linux system equipped with 64 Intel(R) Xeon(R) Gold 6226R CPUs @ 2.90GHz, 376GB of RAM, and a single GeForce RTX 3090 GPU (24GB). The model is implemented in PyTorch~\citep{paszke2019pytorch} and DGL~\citep{wang2019deep}.

\textbf{Data Preparation.} 
Following prior studies~\citep{hu2025modality,zhou2023tale}, we conduct experiments on three public datasets—Baby, Sports, and Clothing—from the Amazon review corpus~\citep{mcauley2013amateurs}. Each dataset contains user–item interactions, item images, and textual descriptions.
Following prior work~\citep{zhou2023tale}, we adopt 4,096-dimensional visual embeddings encoded by Convolutional Neural Networks~\citep{he2016ups} and 384-dimensional text embeddings encoded by sentence-transformers~\citep{reimers2019sentence}.
Following prior practice~\citep{li2025generating}, we simulate missing modalities by randomly masking a proportion of item modalities, leaving $m\in[1, M-1]$ available modalities (where $M=2$ for the evaluated datasets), ensuring that at least one modality remains. Masked modalities are initialized as zero vectors.
The dataset statistics are shown in Table~\ref{tab:stat}.

\textbf{Baselines.} 
To evaluate the effectiveness of GRE-MC, we compare it against two categories of methods:
\textit{(1) General multimodal recommendation models.}  
BPR~\citep{rendle2009bpr},
LightGCN~\citep{he2020lightgcn},
SLMRec~\citep{tao2022self},
FREEDOM~\citep{zhou2023tale},  
BM3~\citep{zhou2023bootstrap},  
DRAGON~\citep{zhou2023enhancing}, 
PGL~\citep{yu2025mind}, and
MIG-GT~\citep{hu2025modality}.
Notably, as these baselines assume full modalities during both training and evaluation, we adopt a simple imputation strategy by replacing missing modality features with zero vectors.
\textit{(2) Missing-modality-aware models.}  
Imputed~\citep{malitesta2024we},
MILK~\citep{bai2024multimodality}, and
MoDiCF~\citep{li2025generating}.
Among them, Imputed and MoDiCF explicitly perform modality completion. For Imputed, we adopt the default downstream recommendation model, i.e., MIG-GT, in our experiments. For MoDiCF, we employ its proposed counterfactual multimodal recommendation module.

\begin{table}[!t]
\centering
\caption{Statistics of the datasets with a missing rate of 40\%. $n_\text{full}$, $n_{v\text{-miss}}$, and $n_{t\text{-miss}}$ denote the numbers of items with full modalities, missing visual modality, and missing textual modality, respectively.}
\resizebox{1\linewidth}{!}{
\begin{tabular}{@{}cccccc@{}}
\toprule
Dataset & Users & Items & $n_\text{full}$ / $n_{v\text{-miss}}$ / $n_{t\text{-miss}}$
 & Interactions & Sparsity \\ 
\midrule
Baby & 19,445 & 7,050 & 1,410 / 2,833 / 2,807 & 160,792 & 99.88\% \\
Sports & 35,598 & 18,357 & 3,672 / 7,403 / 7,282 & 296,337 & 99.95\% \\
Clothing & 39,387 & 23,033 & 4,607 / 9,251 / 9,175 & 278,677 & 99.97\% \\
\bottomrule
\end{tabular}
}
\label{tab:stat}
\end{table}

\begin{table*}[!t]
\centering
\caption{Recommendation performance (\%) across methods (MM.: multimodal; MA.: missing-modality-aware). The best results are shown in bold. The \textit{Improve} row denotes the relative improvement (\%) of our method over the second-best result. An asterisk ($^{*}$) indicates a significant improvement with $p < 0.05$.}
\resizebox{1\linewidth}{!}{
\begin{tabular}{@{}c|c|c|cccc|cccc|cccc@{}}
\toprule
\multirow{2}{*}{Methods} & \multirow{2}{*}{MM.} & \multirow{2}{*}{MA.} & \multicolumn{4}{c|}{Baby} & \multicolumn{4}{c|}{Sports} & \multicolumn{4}{c}{Clothing} \\
 &  &  & R@10 & R@20 & N@10 & N@20 & R@10 & R@20 & N@10 & N@20 & R@10 & R@20 & N@10 & N@20 \\ \midrule
BPR & \xmark & \xmark & \ms{3.57}{0.35} & \ms{5.75}{0.30} & \ms{1.92}{0.27} & \ms{2.49}{1.01} & \ms{4.32}{1.00} & \ms{6.53}{0.34} & \ms{2.41}{0.68} & \ms{2.98}{0.63} & \ms{1.99}{0.95} & \ms{2.98}{0.35} & \ms{1.06}{1.19} & \ms{1.32}{1.19} \\
LightGCN & \xmark & \xmark & \ms{4.79}{0.21} & \ms{7.54}{0.61} & \ms{2.57}{0.69} & \ms{3.28}{0.30} & \ms{5.69}{0.25} & \ms{8.64}{0.03} & \ms{3.11}{0.02} & \ms{3.87}{0.14} & \ms{2.99}{0.69} & \ms{4.49}{0.84} & \ms{1.67}{0.63} & \ms{2.05}{0.16} \\ \midrule
SLMRec & \cmark & \xmark & \ms{5.31}{0.74} & \ms{7.73}{1.09} & \ms{2.93}{0.86} & \ms{3.55}{0.79} & \ms{6.46}{0.54} & \ms{9.74}{0.83} & \ms{3.46}{0.36} & \ms{4.27}{0.11} & \ms{4.38}{0.43} & \ms{6.29}{0.24} & \ms{2.39}{0.07} & \ms{2.87}{0.14} \\
FREEDOM & \cmark & \xmark & \ms{4.60}{0.40} & \ms{7.51}{0.89} & \ms{2.41}{0.90} & \ms{3.15}{0.19} & \ms{5.94}{0.37} & \ms{8.98}{0.16} & \ms{3.26}{0.92} & \ms{4.05}{0.26} & \ms{3.62}{0.26} & \ms{5.37}{0.64} & \ms{2.01}{0.48} & \ms{2.46}{0.36} \\
BM3 & \cmark & \xmark & \ms{4.99}{0.38} & \ms{8.01}{0.20} & \ms{2.62}{0.87} & \ms{3.39}{0.63} & \ms{5.82}{0.09} & \ms{9.06}{0.16} & \ms{3.18}{0.79} & \ms{4.01}{0.28} & \ms{4.38}{0.38} & \ms{6.29}{0.55} & \ms{2.39}{0.45} & \ms{2.87}{0.47} \\
DRAGON & \cmark & \xmark & \ms{5.36}{0.22} & \ms{8.70}{0.74} & \ms{2.89}{0.56} & \ms{3.75}{0.93} & \ms{6.28}{1.18} & \ms{9.34}{0.15} & \ms{3.46}{0.34} & \ms{4.24}{0.27} & \ms{4.40}{0.90} & \ms{6.58}{1.14} & \ms{2.45}{0.44} & \ms{3.00}{0.93} \\
PGL & \cmark & \xmark & \ms{5.15}{0.78} & \ms{8.21}{0.96} & \ms{2.73}{1.13} & \ms{3.52}{0.13} & \ms{5.70}{0.90} & \ms{8.67}{0.86} & \ms{3.12}{0.35} & \ms{3.88}{0.12} & \ms{4.38}{0.86} & \ms{6.29}{0.22} & \ms{2.39}{0.41} & \ms{2.87}{0.36} \\
MIG-GT & \cmark & \xmark & \ms{5.48}{0.83} & \ms{8.65}{1.13} & \ms{2.86}{0.58} & \ms{3.74}{0.83} & \ms{6.31}{1.11} & \ms{9.57}{0.80} & \ms{3.42}{0.02} & \ms{4.27}{0.60} & \ms{4.43}{0.12} & \ms{6.64}{0.42} & \ms{2.42}{0.20} & \ms{2.99}{0.25} \\ \midrule
Imputed & \cmark & \cmark & \ms{5.13}{0.47} & \ms{8.29}{0.75} & \ms{2.76}{0.08} & \ms{3.58}{0.76} & \ms{6.21}{0.15} & \ms{9.68}{0.19} & \ms{3.30}{0.71} & \ms{4.37}{0.31} & \ms{4.49}{0.23} & \ms{6.08}{0.80} & \ms{2.39}{0.14} & \ms{3.01}{0.23} \\
MILK & \cmark & \cmark & \ms{4.81}{0.74} & \ms{8.28}{0.44} & \ms{2.15}{0.52} & \ms{2.97}{0.09} & \ms{6.22}{1.20} & \ms{9.45}{0.08} & \ms{3.12}{0.66} & \ms{4.18}{0.04} & \ms{4.36}{0.42} & \ms{5.75}{0.39} & \ms{2.18}{0.21} & \ms{2.98}{1.16} \\
MoDiCF & \cmark & \cmark & \ms{5.51}{0.17} & \ms{8.76}{0.21} & \ms{2.95}{0.10} & \ms{3.78}{0.10} & \ms{6.51}{0.66} & \ms{9.83}{0.59} & \ms{3.54}{1.04} & \ms{4.42}{1.05} & \ms{4.52}{0.74} & \ms{6.86}{0.36} & \ms{2.50}{0.77} & \ms{3.04}{0.63} \\ \midrule
GRE-MC & \cmark & \cmark & \textbf{\ms{5.84}{0.21}} & \textbf{\ms{9.21}{0.32}} & \textbf{\ms{3.16}{0.30}} & \textbf{\ms{4.03}{0.21}} & \textbf{\ms{6.66}{0.37}} & \textbf{\ms{10.28}{0.31}} & \textbf{\ms{3.70}{0.23}} & \textbf{\ms{4.63}{0.31}} & \textbf{\ms{4.72}{0.43}} & \textbf{\ms{7.15}{0.29}} & \textbf{\ms{2.60}{0.35}} & \textbf{\ms{3.21}{0.24}} \\
Improv. & - & - & *5.99\% & *5.14\% & *7.12\% & *6.61\% & 2.30\% & *4.58\% & *4.52\% & *4.75\% & *4.42\% & *4.23\% & *4.00\% & *5.59\% \\ \bottomrule
\end{tabular}
}
\label{tab:results}
\end{table*}

\textbf{Evaluation Scheme.} 
To ensure a consistent comparison, we follow the evaluation scheme commonly adopted in prior work~\citep{hu2025modality,li2025generating}. We use two standard metrics: Recall (R) and Normalized Discounted Cumulative Gain (NDCG), and report performance at cutoffs $K=10$ and $20$, denoted as R@10, R@20, N@10, and N@20, respectively. 
For completion-only models (Imputed and GRE-MC), we uniformly adopt MIG-GT as the recommendation backbone.
We split the datasets by allocating 80\% of user interactions for training, 10\% for validation, and 10\% for testing. We report the mean and standard deviation over five runs with different random seeds.

\textbf{Hyperparameter Settings.} 
By default, we tune a set of hyperparameters via grid search on the validation set and fix the others following established work. We set the missing rate $r=40\%$ and hidden dimension $d = 128$. For GRE-MC, the number of anchors and codebook size are tuned over \(K\in\{5, 10, 20\}\) and \(C\in\{10, 50, 100\}\), respectively. 
We set the MAGE iteration cap $T=10$ to avoid significant overhead. 
The LPE dimension is fixed at $k=20$ following established work~\citep{dwivedi2023benchmarking}, and we use $L=2$ Transformer layers with $4$ heads following prior work~\citep{zhou2024rethinking}. 
For the sparse-routing codebook, we set the top $P=4$ entry selection, routing temperature \(\tau = 0.5\), and noise standard deviation $0.1$, following prior practices~\citep{jang2017categorical}. 
We tune the regularization weights $\lambda_\mathrm{usage},\lambda_\mathrm{load}\in \{0.5,1.0,2.0\}$.
Models are trained with the Adam optimizer~\citep{kingma2017adammethodstochasticoptimization}, learning rate \(\eta \in \{0.01, 0.003, 0.001, 0.0003, 0.0001\}\), L2 regularization weight $\lambda_\mathrm{L_2} \in \{1e-4, 1e-5, 1e-6\}$, batch size $512$, and dropout rate $0.5$, following common settings~\citep{jin2022graph}.

\subsection{Performance Analysis}

From Table~\ref{tab:results}, several clear trends emerge across datasets and metrics. We summarize the key observations as follows:

\begin{itemize}
\item Structure-based models (BPR and LightGCN) rely solely on user–item interactions without incorporating multimodal features and thus underperform across all datasets. Their limited representational capacity underscores the importance of leveraging multimodal data for recommendation.
\item General multimodal recommendation models that are not specifically designed to handle missing modalities (e.g., SLMRec, FREEDOM) consistently exhibit suboptimal performance. This highlights the limitations of traditional multimodal fusion methods, which assume fully observed modalities. Their degraded performance emphasizes the necessity of addressing missing modalities.
\item Missing-modality-aware methods (e.g., Imputed and MoDiCF) demonstrate strong performance even with incomplete features. These methods effectively manage missing modalities, maintaining stable and high-quality recommendation performance under missing-modality scenarios.
\item GRE-MC consistently achieves SOTA results with 2.3\%–7.1\% improvements over the strongest baselines. Unlike methods based on self information (e.g., MoDiCF), our approach integrates graph semantics and structure for more informed completion. Compared with neighbor-based methods (e.g., Imputed), GRE-MC employs modality-aware graph retrieval to construct more relevant subgraphs, leading to higher-quality completion and robust recommendations.
\end{itemize}

\subsection{Detailed Analysis}

\subsubsection{Ablation Study}
\begin{figure}[!t]
    \centering
    \subfloat[R@10]{
        \includegraphics[width=0.48\linewidth]{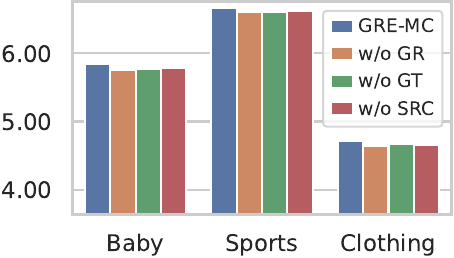}
    }\hfill
    \subfloat[R@20]{
        \includegraphics[width=0.48\linewidth]{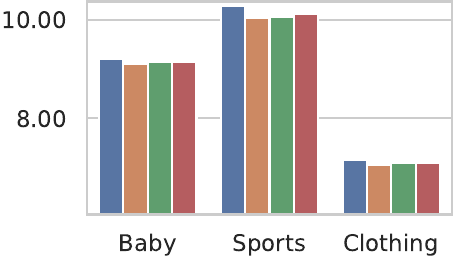}
    }\\
    \subfloat[N@10]{
        \includegraphics[width=0.48\linewidth]{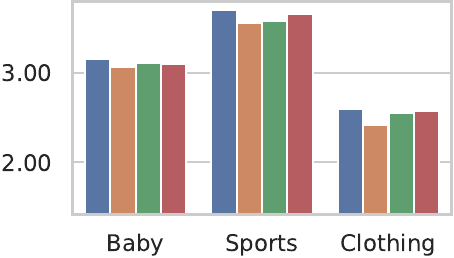}
    }\hfill
    \subfloat[N@20]{
        \includegraphics[width=0.48\linewidth]{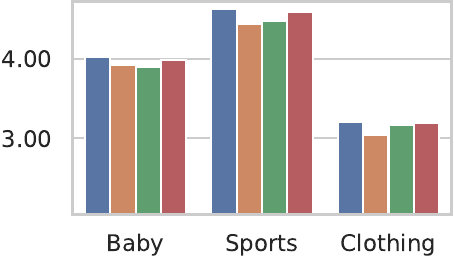}
    }
    \caption{Ablation study of the main components: modality-aware graph retrieval (GR), graph transformer (GT), and sparse-routing codebook (SRC).}
    \label{fig:ablation}
\end{figure}

Figure~\ref{fig:ablation} presents ablation results that highlight the contribution of each core component in GRE-MC. Removing the retrieval module generally results in the most significant performance drop. This supports our intuition that local aggregation fails under limited observations and that subgraph expansion from semantically similar nodes plays a pivotal role in completion. 
Removing the graph transformer or the sparse-routing codebook also leads to performance degradation, albeit more moderately. 
The graph transformer plays a crucial role for integrating semantic and structural information from the retrieved subgraph.
Although the sparse-routing codebook provides a relatively small contribution to the final performance, it effectively captures high-level multimodal semantics with fewer parameters, contributing to both generalization and performance.

\begin{figure}[!t]
    \centering
    \subfloat[Baby R@20\label{fig:baby_r20}]{
        \includegraphics[width=0.48\linewidth]{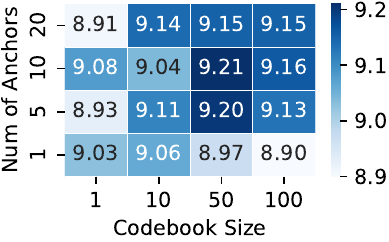}
    }\hfill
    \subfloat[Baby N@20\label{fig:baby_n20}]{
        \includegraphics[width=0.48\linewidth]{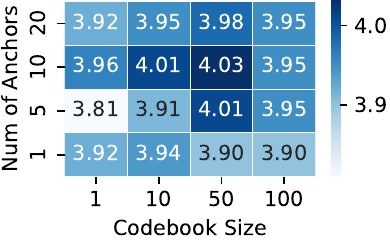}
    }\\
    \subfloat[Clothing R@20\label{fig:clothing_r20}]{
        \includegraphics[width=0.48\linewidth]{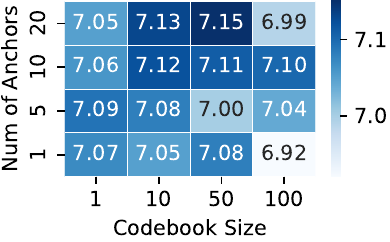}
    }\hfill
    \subfloat[Clothing N@20\label{fig:clothing_n20}]{
        \includegraphics[width=0.48\linewidth]{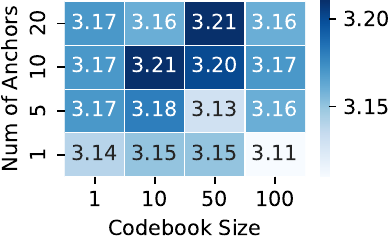}
    }
    \caption{Impact of the number of anchors and codebook size.}
    \label{fig:anchors-codes}
\end{figure}

\subsubsection{Impact of Number of Anchors and Codebook Size}
In Figure~\ref{fig:anchors-codes}, we examine the impact of the number of anchors $K$ and the codebook size $C$. 
Increasing the number of anchors initially leads to noticeable performance gains, but the improvement plateaus and may slightly decline when too many anchors are used, likely due to the inclusion of noisy neighbors. Similarly, the codebook size exhibits a non-linear effect: both overly small and overly large codebooks underperform, while moderate sizes strike an effective balance. Overall, using a moderate number of anchors and a compact codebook yields strong results with minimal overhead.

\begin{figure}[!t]
\centering
\subfloat[R@10]{
    \includegraphics[width=0.48\linewidth]{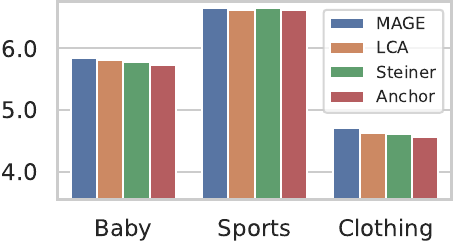}
}\hfill
\subfloat[R@20]{
    \includegraphics[width=0.48\linewidth]{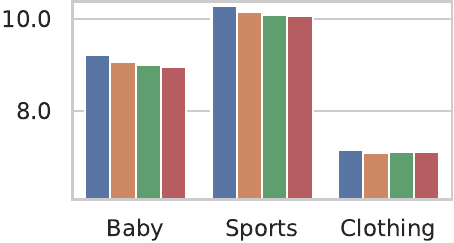}
}\\
\subfloat[N@10]{
    \includegraphics[width=0.48\linewidth]{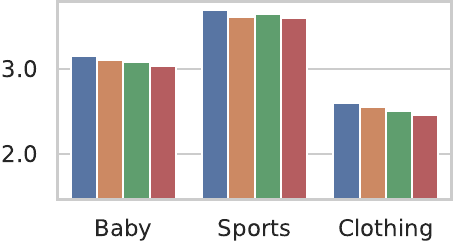}
}\hfill
\subfloat[N@20]{
    \includegraphics[width=0.48\linewidth]{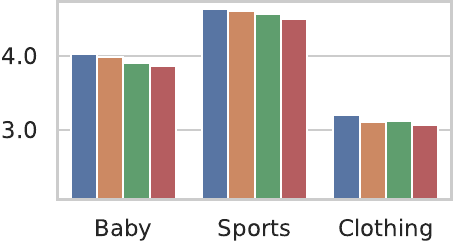}
}
\caption{Impact of different graph retrieval methods. \textit{Anchor} refers to the setting without graph retrieval.}
\label{fig:retriever}
\end{figure}

\subsubsection{Impact of Graph Retrieval Methods}
We compare graph retrieval strategies: MAGE (Algorithm~\ref{alg:mage}), ACS (Algorithm~\ref{alg:acs}), the Steiner tree implemented by NetworkX~\citep{hagberg2008exploring}, and the anchor-only setting. As shown in Figure~\ref{fig:retriever}, MAGE achieves the best performance, with ACS and the Steiner tree offering improvements over the anchor-only setting. 
Both ACS and the Steiner tree focus on constructing small connected subgraphs that link the anchor nodes, which makes them effective at preserving local connectivity but limits their ability to capture richer contextual signals. In contrast, MAGE extends ACS by iteratively retrieving nodes with higher semantic relevance and removing less relevant nodes, thereby enriching the subgraph with more informative features. This enhanced graph context leads to better modality completion quality, accounting for MAGE’s consistently superior performance.

\subsubsection{Performance under Different Missing Rates}
\begin{figure}[!t]
    \centering
    \subfloat[Baby N@20]{
        \includegraphics[width=0.31\linewidth]{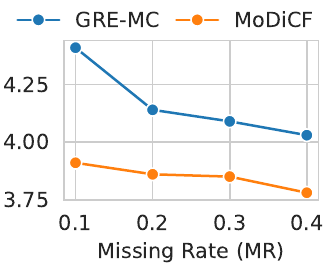}
    }\hfill
    \subfloat[Sports N@20]{
        \includegraphics[width=0.31\linewidth]{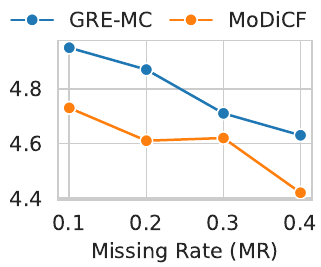}
    }\hfill
    \subfloat[Clothing N@20]{
        \includegraphics[width=0.31\linewidth]{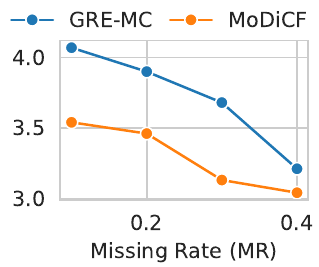}
    }
    \caption{Performance under different missing rates.}
    \label{fig:mr}
\end{figure}

We evaluate performance on three datasets under different missing rates using the N@20 metric. As illustrated in Figure~\ref{fig:mr}, GRE-MC consistently surpasses the baseline, demonstrating both effectiveness and robustness in handling diverse incomplete scenarios.

\subsubsection{Relevance Comparison between Neighbor and Retrieved Subgraphs}
\begin{figure}[!t]
    \centering
    \subfloat[Baby (image missing)]{
        \includegraphics[width=0.48\linewidth]{figures/baby_relevance_image.pdf}
    }\hfill
    \subfloat[Baby (text missing)]{
        \includegraphics[width=0.48\linewidth]{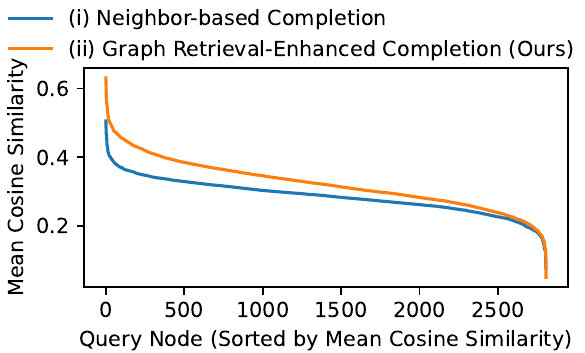}
    }\\
    \subfloat[Sports (image missing)]{
        \includegraphics[width=0.48\linewidth]{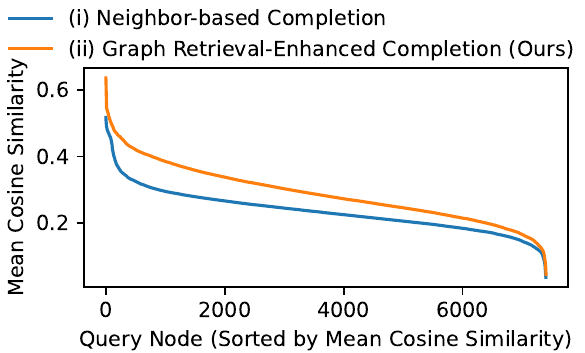}
    }\hfill
    \subfloat[Sports (text missing)]{
        \includegraphics[width=0.48\linewidth]{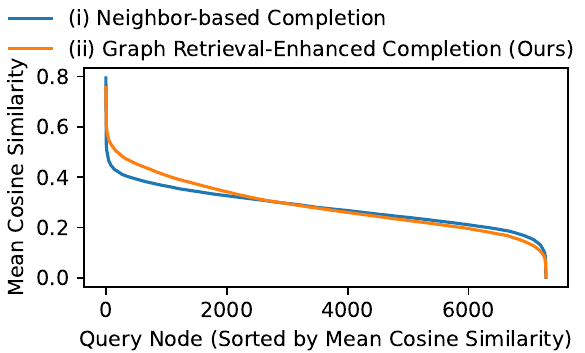}
    }
    \caption{Relevance comparison between neighbor and retrieved subgraphs.}
    \label{fig:relevances}
\end{figure}

Figure~\ref{fig:relevances} compares the subgraphs either obtained from a query node's neighborhood or its retrieved subgraph.
When a modality is missing, neighbor subgraphs often contain items that are structurally close but semantically less informative.
In contrast, retrieved subgraphs bring in globally relevant items with similar observed modalities, providing more informative context for completion.
This observation motivates our graph retrieval-enhanced approach to modality completion.

\begin{table}[!t]
\caption{Runtime (seconds) across different stages.}
\resizebox{0.7\linewidth}{!}{
\begin{tabular}{@{}ccrrr@{}}
\toprule
Method & Stage & Baby & Sports & Clothing \\ \midrule
\multirow{4}{*}{GRE-MC} & Graph Retrieval & 22 & 119 & 162 \\
 & Completion & 880 & 1,269 & 1,601 \\
 & Recommendation & 200 & 473 & 1,015 \\
\cmidrule(lr){2-5}
 & Total & \textbf{1,102} & \textbf{1,861} & \textbf{2,778} \\ \midrule
\multirow{3}{*}{MoDiCF} & Completion & 1,188 & 2,826 & 4,280 \\
 & Recommendation & 14,996 & 15,352 & 16,193 \\
\cmidrule(lr){2-5}
 & Total & 16,184 & 18,178 & 20,473 \\ \bottomrule
\end{tabular}
}
\label{tab:time}
\end{table}

\subsubsection{Runtime and Scalability}
Table~\ref{tab:time} reports the runtime breakdown across different stages.
We observe that graph retrieval is consistently efficient, incurring only negligible overhead, which shows that the worst-case complexity of $O(|E|)$ for MAGE is not a practical bottleneck in real-world graphs.
In contrast, modality completion introduces an additional cost of less than 800 seconds compared to the downstream recommendation stage. This overhead increases only mildly with graph size, indicating good scalability.
Notably, compared with MoDiCF, our GRE-MC incurs significantly lower computational costs in both completion and recommendation, as MoDiCF relies on computationally intensive modules such as diffusion-based completion and counterfactual recommendation.

\subsubsection{Impact of Codebook Regularization}\label{ap:reg}

\begin{figure}[t]
  \centering
  \subfloat[N@10\label{fig:usage-n10}]{
    \includegraphics[width=0.48\linewidth]{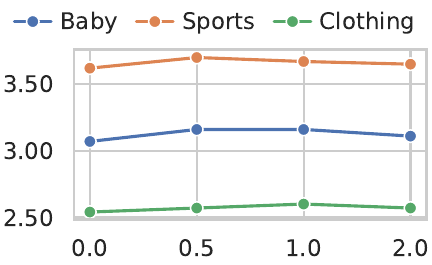}
  }\hfill
  \subfloat[N@20\label{fig:usage-n20}]{
    \includegraphics[width=0.48\linewidth]{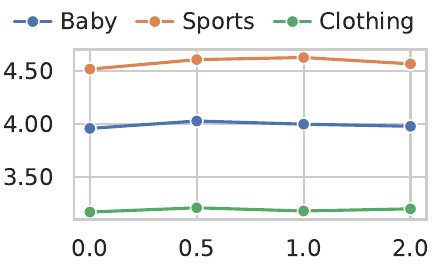}
  }\\
  \subfloat[R@10\label{fig:usage-r10}]{
    \includegraphics[width=0.48\linewidth]{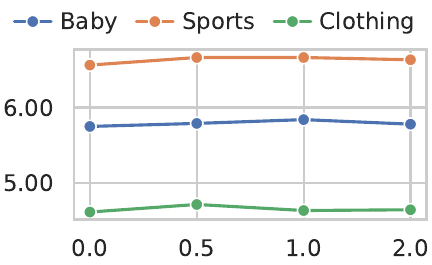}
  }\hfill
  \subfloat[R@20\label{fig:usage-r20}]{
    \includegraphics[width=0.48\linewidth]{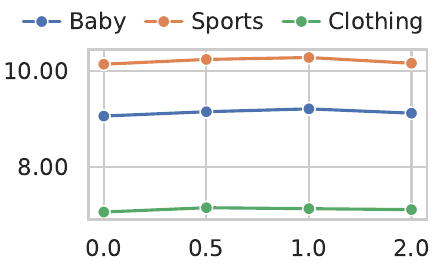}
  }
  \caption{Impact of codebook soft usage weight $\lambda_\mathrm{usage}$.}
  \label{fig:lambda-usage}
\end{figure}

\begin{figure}[t]
  \centering
  \subfloat[N@10\label{fig:load-n10}]{
    \includegraphics[width=0.48\linewidth]{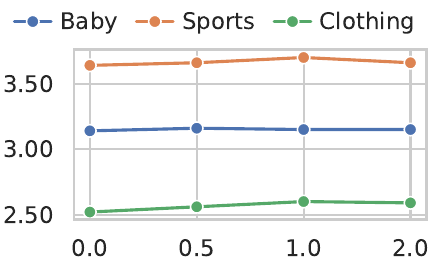}
  }\hfill
  \subfloat[N@20\label{fig:load-n20}]{
    \includegraphics[width=0.48\linewidth]{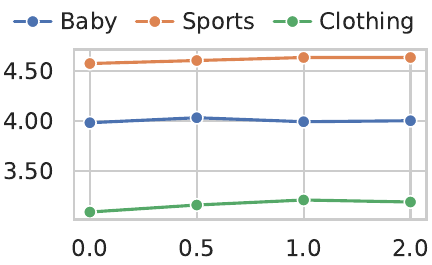}
  }\\
  \subfloat[R@10\label{fig:load-r10}]{
    \includegraphics[width=0.48\linewidth]{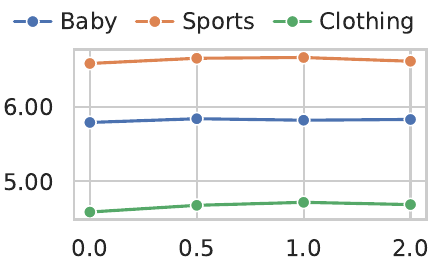}
  }\hfill
  \subfloat[R@20\label{fig:load-r20}]{
    \includegraphics[width=0.48\linewidth]{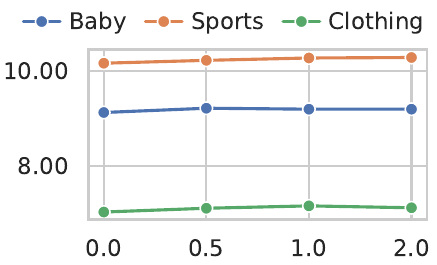}
  }
  \caption{Impact of codebook load balancing weight $\lambda_\mathrm{load}$.}
  \label{fig:lambda-load}
\end{figure}

We analyze how the two regularization terms, i.e., $\mathcal{L}_{\mathrm{usage}}$ and $\mathcal{L}_{\mathrm{load}}$, influence the final recommendation performance.  
As shown in Figs.~\ref{fig:lambda-usage} and \ref{fig:lambda-load}, both usage and load-balancing regularization consistently improve performance compared to the model without the corresponding regularization terms.  
Moreover, the results remain smooth and stable across a wide range of weight values, indicating that the model is relatively insensitive to these hyperparameters.

\section{Conclusion}
In this paper, we proposed GRE-MC, a graph retrieval–enhanced framework for modality completion. GRE-MC introduces a modality-aware subgraph retrieval mechanism that selects semantically relevant subgraphs to provide richer contextual signals for reconstructing missing modalities. A graph transformer is further employed to jointly encode the query node and retrieved subgraph for modality completion, while a sparse-routing codebook regularizes latent representations for improved robustness. Extensive experiments on benchmark datasets demonstrate that GRE-MC consistently outperforms SOTA approaches, validating the effectiveness of retrieval-enhanced and codebook-regularized modeling for multimodal completion.

\section*{Acknowledgments}
This research is supported by the National Research Foundation, Singapore and Infocomm Media Development Authority under its Trust Tech Funding Initiative, and Ministry of Education AcRF Tier 1 grant (No. T1 251RES2315) in Singapore. Any opinions, findings and conclusions or recommendations expressed in this material are those of the author(s) and do not reflect the views of National Research Foundation, Singapore and Infocomm Media Development Authority.

\bibliographystyle{ACM-Reference-Format}
\balance
\bibliography{sample-base}

\end{document}